# ANT COLONY OPTIMIZATION WITH A NEW RANDOM WALK MODEL FOR COMMUNITY DETECTION IN COMPLEX NETWORKS

DI JIN, DAYOU LIU, BO YANG, JIE LIU*, DONGXIAO HE

*College of Computer Science and Technology, Jilin University*
*Changchun, 130012, China*
*jindi.jlu@gmail.com*
*liudy@jlu.edu.cn*
*ybo@jlu.edu.cn*
*liu_jie@jlu.edu.cn*
*hedongxiaojlu@gmail.com*

Detecting communities from complex networks has recently triggered great interest. Aiming at this problem, a new ant colony optimization strategy building on the Markov random walks theory, which is named as MACO, is proposed in this paper. The framework of ant colony optimization is taken as the basic framework in this algorithm. In each iteration, a Markov random walk model is employed as heuristic rule; all of the ants' local solutions are aggregated to a global one through an idea of clustering ensemble, which then will be used to update a pheromone matrix. The strategy relies on the progressive strengthening of within-community links and the weakening of between-community links. Gradually this converges to a solution where the underlying community structure of the complex network will become clearly visible. The proposed MACO has been evaluated both on synthetic benchmarks and on some real-world networks, and compared with some present competing algorithms. Experimental result has shown that MACO is highly effective for discovering communities.

## 1. Introduction

Many complex systems in the real world exist in the form of networks, such as social networks, biological networks, Web networks, etc., which are also often classified as complex networks. Complex network analysis has been one of the most popular research areas in recent years due to its applicability to a wide range of disciplines [1, 2, 3]. While a considerable body of work addressed basic statistical properties of complex networks, such as the existence of "small world effect" [1] and the presence of "power laws" in the link distribution [2], another property has also been paid particular attention, that is, "community structure": the nodes in networks are often found to cluster into tightly-knit groups with a high density of within-group edges and a lower density of between-group edges [3]. The community detection problem (CDP), which is also a network clustering task, is to detect and interpret community structures from various complex network data sets.

The research on community detection in complex networks is of fundamental importance. It has both theoretical significance and practical applications in terms of analyzing network topology, comprehending network function, unfolding network patterns and forecasting network activities. It has been used in many areas, such as terrorist organization recognition, organization management, biological network analysis, Web community mining, topic based Web document clustering, search engine, link prediction, etc [4].

So far, lots of community detection algorithms have been developed. In terms of the basic strategies adopted by them, they mainly fall into two main categories: optimization and heuristic based methods. The former solves the CDP by transforming it into an optimization problem and trying to find an optimal solution for a predefined objective function such as the network modularity employed in several algorithms [5, 6, 7, 8, 9]. In contrast, there are no explicit optimization objectives in the heuristic based methods, and they solve the CDP based on some intuitive assumptions or heuristic rules, such as in the Girvan-Newman (GN) algorithm [3], Clique Percolation Method (CPM) [10], Label Propagation Algorithm (LPA) [11], Community Detection with Propinquity Dynamics (CDPD) [12], Opinion Dynamics with Decaying Confidence (ODDC) [13], etc.

As a kind of special heuristic strategy, Markov theory based random walk models have also been widely used in this area. For CDP, Dongen [14] proposed Markov Cluster algorithm (MCL) which is based on random walks on a graph and uses simple algebraic operations on its associated stochastic

---

matrix. Pons et al. [15] proposed a measure of similarities between network vertices based on random walks and developed an agglomerative algorithm to efficiently compute the community structure of networks by using this measure. Gunes et al. [16] presented an agent-based community detection algorithm by making the agents establish a random walk on a network. Rosvall et al. [17] used the probability flow of random walks on a network as a proxy for information flows in the real system and decomposed the network into modules by compressing a description of the probability flow. Weinan E et al. [18] proposed a strategy along the lines of optimal prediction for the Markov chains associated with the dynamics on these networks and developed the necessary ingredients for such an optimal partition strategy. For community detection in signed networks, Yang et al. proposed a heuristic signed networks clustering algorithm (FEC) based on Markov random walk model [19], and then they found that the dynamics of such a stochastic model naturally reflects the intrinsic properties of networks with community structures, and proposed a spectral method to detect network community structure based on large deviation theory [20]. Algorithm MACO proposed in this paper also belongs to this type of algorithms.

Though there are lots of community detection algorithms which have been presented recently, how to further improve the performance is still an open problem. In order to address this problem, a Markov random walk theory based ant colony optimization is proposed in this paper enlightened by Ref. 19. In the algorithm, each ant detects its own community by using a new random walk model as heuristic rule; in each iteration, all the ants collectively produce the current solution via the thought of clustering ensemble [21], and update their pheromone matrix by using this solution; at last, after the algorithm has converged, the pheromone matrix is analyzed in order to attain the community structure for the target network.

## 2. Algorithm

### 2.1. *The main idea*

Let $N = (V, E)$ denote an unweighted and undirected network, where $V$ is the set of nodes (or vertices) and $E$ is the set of edges (or links). Let a $k$-way partition of the network be defined as $\pi = \{V_1, V_2, \ldots, V_k\}$, where $V_1, V_2, \ldots, V_k$ satisfy $\bigcup_{1 \leq i \leq k} V_i = V$ and $\bigcap_{1 \leq i \leq k} V_i = \varnothing$. If partition $\pi$ has the property that within-community edges are dense and between-community edges are sparse, it's called a well defined community structure of this network.

In view of a stochastic process defined on $N$, in which an imaginary agent freely walks from one node to another along the links between them. When the agent arrives at one node, it will randomly select one of its neighbors and move there. Considering the inherent community property of the network, the random walk agent should be found it difficult to move outside its own community boundary, whereas it should be easy for the agent to reach other nodes within its community, as link density within a community should be high, by definition. In other words, the probability of remaining in the same community, that is, an agent starts from any node and stays in its own community after freely walking by a number of steps, should be greater than that of going out to a different community.

Based on the above general observation, in our ant colony optimization strategy, each ant (only different from agent in the sense that it can consult and update a "pheromone" variable in each link) takes random walk as heuristic rule, and meanwhile, is directed by pheromone to find its solution. In each iteration, the solution found by each ant only expresses its local view, while a global solution will be attained after aggregating all of the ants' local solutions to one through a thought of clustering ensemble [21], which then will be used to update the pheromone matrix. As the process evolves, the community characteristic of the pheromone matrix will gradually become sharper and our algorithm MACO converges to a solution where the community structure can be accurately detected. In short, the pheromone matrix can be regarded as the final clustering result which aggregates the information of all the ants in all iterations within this algorithm. Thus, as we can see, this algorithm may share some common features with the basic thought of some clustering ensemble methods.

In order to further clarify the above idea, an intuitive description is presented as follows. Assume a network $N$ with obvious community structure, in which some ants freely crawl along the links. The ants have a given life-cycle, and the new ant colony will be generated immediately when all of the former ants die. At the beginning of this algorithm, there is yet no impact of the pheromone on network $N$. Only due to the restriction by community structure, the ant's probability of remaining in its own community should be greater than that of going out to other communities, but there is no difference between these ants and the random walk agents at the moment since pheromone distribution is still homogeneous. As ants move, with the accumulation and volatilization of pheromone left by the former ants, the pheromone on within-community links will become thicker and thicker, and the pheromone on between-community links will become thinner and thinner. In fact, pheromone is simply a mechanism that can register past walks in the network and that leads to more informed decisions for subsequent walks. The process strengthens the trend that any ant will more often remain in its own community. At last, when the pheromone matrix converges, the community structure of network $N$ will be attained naturally. To sum up, the core of our MACO is that, by strengthening within-community links and weakening between-community links, an underlying community structure of the network will gradually become visible.

## 2.2. *A local solution by one ant*

In our ant colony optimization framework, guided by heuristic rule and pheromone effect, each ant needs to produce its local solution which is actually the community that it is situated. This method includes two parts. The goal of the first part is to unfold the ant's community by calculating its $l$-step transition probability distribution; and the goal of the second one is to extract the emerged community by designing a suitable cut strategy.

### 2.2.1. *Unfold the ant's community*

Assume that the adjacency matrix of network $N$ is $A = (a_{ij})_{n \times n}$. When we consider the current pheromone matrix $B = (b_{ij})_{n \times n}$ of the ants, the original unweighted network $N$ will then become a pheromone weighted network $W$, whose adjacency matrix is $M = (m_{ij})_{n \times n} = (a_{ij} \cdot b_{ij})_{n \times n}$. Thus, we only consider network $W$ in this section.

Let an ant freely crawls on network $W$. Assume that $X = \{X_t, t \geq 0\}$ denote the ant's positions, and $P\{X_t = i, 1 \leq i \leq n\}$ denote the probability that the ant arrives at node $i$ after $t$ steps walking. For $i_t \in V$ we have $P\{X_t = i_t \mid X_0 = i_0, X_1 = i_1, \ldots, X_{t-1} = i_{t-1}\} = P\{X_t = i_t \mid X_{t-1} = i_{t-1}\}$. That is, the next state of the ant is completely decided by its previous state, which is called a Markov property. So, this stochastic process is a discrete Markov chain and its state space is the node set $V$. Furthermore, $X_t$ is homogeneous because $P\{X_t = j \mid X_{t-1} = i\} = p_{ij}$, where $p_{ij}$ is the transition probability from node $i$ to node $j$ on network $W$. Then the $p_{ij}$ is defined as Eq. (1).

$$p_{ij} = \frac{m_{ij}}{\sum_r m_{ir}} \tag{1}$$

Let we consider the above Markov model. Given a specific source node $s$ for an ant, let $\alpha_s^l(i)$ denotes the probability that this ant starts from node $s$ and eventually arrives at an arbitrary destination node $i$ within $l$ steps. The value of $\alpha_s^l(i)$ can be estimated iteratively by Eq. (2).

$$\alpha_s^l(i) = \sum_{r=1}^n \alpha_s^{l-1}(r) \cdot p_{ri} \tag{2}$$

Here $\alpha_s^l$ is called the $l$ step transition probability distribution (vector). Note that the sum of the probability values arriving at all the nodes from source node $s$ will be 1, that is $\sum_{i=1}^n \alpha_s^l(i) = 1$. When step number $l$ equals to 0, which means the ant still stays on node $s$, then $\alpha_s^0(s)$ equals to 1 and $\alpha_s^0(i)$ equals to 0 for each $i \neq s$.

As the link density within a community is, in general, much higher than that between communities, an ant that starts from the source node $s$ should have more paths to choose from to reach the nodes in its own community within $l$ steps, when the value of $l$ is suitable. On the contrary, the ant should have much lower probability to arrive at the nodes outside its community. In other words, it will be hard for an ant to fall on a community by passing those "bottleneck" links and to leave the existing community. Thus, in general, vector $\alpha_s^l$ should meet Eq. (3) well when step number $l$ is suitable. In this equation, $C_s$ denotes the community where node $s$ is situated.

$$\forall_{i \in C_s} \forall_{j \notin C_s} : \alpha_s^l(i) > \alpha_s^l(j) \tag{3}$$

However, from the experiment we find out that, though the above Markov method is well suitable for some simple networks such as the Newman benchmark [3] and some small real networks, it is not so effective for some complicated networks, like the Lancichinetti benchmark [22] and some large real networks. This also means $\alpha_s^l$ can't meet Eq. (3) well enough for those relatively complicated networks. Furthermore, this method is very sensitive to the choice of the step number $l$, which will give a crucial effect in its performance.

In order to overcome these drawbacks, a new Markov random walk method combined with a constraint strategy based on the annealed network theory [23] is proposed here. The idea of our method arises in the intuition that a Markov random process on a network with community structure is different from that on its corresponding annealed network without communities. Considering this thought, in each step, the probability that an ant starts from a specific source node $s$ and arrives at each destination node $i$ will be defined as the difference between its associated probability computed on the community network $W$ and that on the corresponding annealed network $R$. Due to $R$ having no community structure, the link density within a community in network $W$ should be much higher than that in $R$, while the link density between communities in $W$ should be much lower than that in $R$. Thus, under the constraint brought by the annealed network, this ant can hardly escape from its associated community and reach the nodes outside. This will cause that, the computed probability value of each within community node will be high, while that of each outside node will be relatively low and most of them even equal to 0 in general. Moreover, its performance is not so sensitive as before to the parameter $l$. Later, we will offer some detailed analysis on this parameter.

Given the network $W = (V', E')$ with its degree distribution $D'$, and its corresponding annealed network $R = (V^*, E^*)$ with its degree distribution $D^*$, there should be $V' = V^*$, $D' = D^*$ and $E' \neq E^*$, which means $W$ and $R$ have the same degree distribution [23]. Let $M = (m_{ij})_{n \times n}$ denote the adjacency matrix of network $W$. There will be $D' = diag(d_1', \ldots d_n')$, in which $d_i' = \sum_j m_{ij}$ denotes the degree of node $i$ in network $W$. Assume that $C = (c_{ij})_{n \times n}$ is the adjacency matrix of $R$. There will be $c_{ij} = d_i' d_j' / \sum_{r=1}^n d_r'$, which denotes the expected number of links between nodes $i$ and $j$ without regard for any community structure [23]. Let $q_{ij}$ denote the transition probability from node $i$ to node $j$ on graph $R$. Thus, it will be defined as Eq. (4).

$$q_{ij} = \frac{c_{ij}}{\sum_r c_{ir}} \quad (4)$$

Considering the constraint generated by this annealed network $R$, let $\beta_s^l(i)$ denotes the probability that this ant starts from the source node $s$ and eventually arrives at an arbitrary destination node $i$ within $l$ steps. The value of $\beta_s^l(i)$ can be estimated iteratively by Eq. (5).

$$\beta_s^l(i) = \max\left(\sum_{r=1}^n \beta_s^{l-1}(r) \cdot p_{ri} - \sum_{r=1}^n \beta_s^{l-1}(r) \cdot q_{ri}, 0\right)$$
$$\beta_s^l(i) = \frac{\beta_s^l(i)}{\sum_{r=1}^n \beta_s^l(r)} \quad (5)$$

It's obvious that $\sum_{r=1}^n \beta_s^{l-1}(r) \cdot p_{ri}$ denotes the transition probability from node $s$ to node $i$ within $l$ steps on network $W$, while $\sum_{r=1}^n \beta_s^{l-1}(r) \cdot q_{ri}$ denotes that probability computed on annealed network $R$. Furthermore, as negative probability is illogical, we make each $\beta_s^l(i)$ is always a nonnegative value. Since the sum of the probability values arriving at all the nodes from source node $s$ should be 1, we normalize $\beta_s^l(i)$ after each step.

Furthermore, as most complex networks have power-law degree distribution, which means there are more paths arriving at the nodes with high degrees than those with low degrees. This will give some negative effects when unfolding communities. Thus, we take into account the effect of power-law degree distribution in complex networks, and propose a further improved $l$ step transition probability distribution $\psi_s^l$ as defined as (6), where $d_i'$ denotes the degree of node $i$ on network $W$. Note that this equation is not iteratively computed as before.

$$\psi_s^l(i) = \frac{\beta_s^l(i)}{d_i'}, \quad \psi_s^l(i) = \frac{\psi_s^l(i)}{\sum_{r=1}^n \psi_s^l(r)} \quad (6)$$

Based on this above idea, the method for an ant (denoted by a source node *s*) to unfold its community can be described as follows. This method is called UC (unfolding community) in short.

S1. Calculate the *l* step transition probability vector of this ant, which is $\psi_s^l$;

S2. Rank all the nodes according to their probability values in descending order, which produces the sorted node list *L*.

S3. Remove the nodes whose associated probability is 0 from *L*;

After these three steps, almost all the within community nodes will be ranked on the top of the sorted node sequence *L*, meanwhile, most of the outside community nodes (probability values are 0) are removed from *L*. Thus, the ant's community has now clearly emerged and is ready for detection. Now, by properly setting a cutoff point (to be explained in the next section), we can precisely extract the ant's community.

**Proposition 1.** The time complexity of the UC is $O(ln^2)$, where *l* is the step number of the ant and *n* is the numbers of nodes.

*Proof.* In S1, It's obvious that the time to compute $\psi_s^l$ by (5) and (6) will be $O(ln^2)$. In S2, the time to rank $\psi_s^l$ will be $O(n\log n)$ in terms of some quick sorting algorithms. In S3, the time is less than $O(n)$. Thus, the time complexity of the UC will be $O(ln^2)$ at last.

2.2.2. *Extract the ant's community*

The ant's community has been emerged from UC by giving a sorted node list *L*, in which almost all the within community nodes have been ranked on the top of *L* and most of the outside community nodes have been removed from *L*. Here we will design a suitable cutoff criterion to extract the ant's community. As any cut strategy should only make use of the original network structure, we just consider the original network $N = (V, E)$ in this section.

In order to propose an effective cutoff method, a well known conductance function [24] which corresponds to the so-called weak definition of community [25] is used here. The conductance can be simply thought of as the ratio between the number of edges inside the community and those leaving it. Let the degree distribution of network *N* is $D = diag(d_1, \ldots d_n)$. Conductance $\phi(S)$ of a set of nodes *S* ($S \subset V$) can be defined as $\phi(S) = c_S/\min(\text{Vol}(S), \text{Vol}(V \setminus S))$, where $c_S$ denotes the size of the edge boundary, $c_S = |\{(u, v) : u \in S, v \notin S\}|$, and $\text{Vol}(S) = \sum_{u \in S} d_u$, where $d_u$ is the degree of node *u* in network *N*. Thus, more community-like sets of nodes have lower conductance. Moreover, this community function has some local characteristics, which is right suitable to extract the emerged community in this section.

Based on the ranked node list *L* obtained from UC, here the emerged community can be easily distilled by finding the cut position which corresponds to the minimum conductance value, and taking it as the cutoff point along this ranked list of nodes. At last, the method to extract the emerged community is summarized as follows. This method is called EC (extract community) in short.

S1. Compute the conductance value of the community corresponding to each cut position of *L*;

S2. Take the community corresponding to the minimum conductance as the extracted one.

**Proposition 2.** The time complexity of the EC is smaller than $O(dn^2)$, where *d* is the average degree of all the nodes in network *N*.

*Proof.* It's obvious that S1 is the most computationally costly step in EC. Let we employ the incremental method to calculate the conductance value for each cut pos in the nodes list *L*. When the cut pos equals to 1, there is $S^1 = \{L(1)\}$ where $L(1)$ denotes the first node in the sorted nodes list *L*, $c_S^1 = d_{L(1)}$, $\text{Vol}(S^1) = d_{L(1)}$, and $\text{Vol}(V \setminus S^1) = m - \text{Vol}(S^1)$ where *m* is the number of edges in the network. When the cut pos equals to *k*, there should be $S^k = S^{k-1} \cup \{L(k)\}$, $c_S^k = c_S^{k-1} + d_{L(k)} - 2 \cdot |S^k \cap N_{L(k)}|$ where $N_{L(k)}$ denotes the neighbors set of node $L(k)$, $\text{Vol}(S^k) = \text{Vol}(S^{k-1}) + d_{L(k)}$, and $\text{Vol}(V \setminus S^k) = m - \text{Vol}(S^k)$. It's obvious that, for each cut pos *k*, $S^k \cap N_{L(k)}$ is the most computationally costly step, whose time complexity is $|S^k| \cdot |N_{L(k)}| = k \cdot d_{L(k)}$. Due to *k* is in the range of $1 \leq k \leq k_{max}$ where $k_{max} \ll n$, thus the time complexity of EC should be $\sum_{k=1}^{k_{max}} k \cdot d_{L(k)}$, which is much smaller than $O(dn^2)$.

Please refer to "*Appendix A*" in order to observe the process in which an arbitrary ant produces its solution at each generation. It's obvious that, in each iteration, the UC subroutine can clearly unfold the ant's local community, meanwhile, the EC subroutine is effective to extract this community.

### 2.3. Algorithm MACO

After offering the sub-method used by each ant to produce its local solution, here we present the complete MACO algorithm which is described as two parts. The first part is called exploration phase. It employs the framework of ant colony optimization so as to finally produce the converged pheromone matrix, which can clearly take on the community feature of the targeted network. The second part is called partition phase, which adopts a simple cut strategy on the converged pheromone matrix to attain the partition of this network.

The exploration phase algorithm (EPA) of the first part is given as follows. Note that this method is described by using the format of Matlab pseudocode.

Procedure EPA
Input: $A$, $T$, $S$, $l$, $\rho$ /* A is the adjacent matrix of the network N, T is the limitation of iteration number, S is the size of ant colony, l is the step number of each ant, ρ is the updating rate of pheromone matrix */
Output: $B$ /* denotes the pheromone matrix */
Begin
1  $B \leftarrow ones(n, n)*n$; /* initialize the pheromone matrix */
2  For $i=1: T$
3   $solution \leftarrow zeros(n, n)$; /* initialize the global solution */
4   For $j=1: S$
5    $s \leftarrow rand(n)$; /* generate a random number s as source node*/
6    $com \leftarrow \mathbf{one\_ant}(A, B, s, l)$; /* generate a local solution (community) by one ant */
7    $solution(com, com) \leftarrow solution(com, com) + 1$;
8   End /* aggregate local solutions of all ants to a global one */
9   $B \leftarrow \rho*B + solution$; /* update the pheromone matrix by using the global solution*/
10  End
End

As we can see from EPA, the framework of ant colony optimization is taken as the basic algorithm framework. In each iteration, each ant detects its local solution (or called its community) via the Markov method proposed in Sec 2.2 (step 6); then the local solutions of all the ants are aggregated to a global one through the thought of clustering ensemble [21] (step 7); and then the global solution is used to update pheromone matrix (step 9). This process stops until the algorithm converges at last.

At the beginning of this algorithm, as no or very little pheromone is left, the solutions of all ants are produced mainly due to the restriction of community structure, and then they will be used to update pheromone matrix. This will make the guiding role of pheromone matrix better targeted, which allows the following ants to produce better solutions. With the increase of iterations, the pheromone matrix is gradually evolving, which makes the ants more and more directed, and the trend that any ant stays in its own community more and more obvious. When the algorithm finally converges, the pheromone matrix can be regarded as the final clustering result which aggregates the information of all the ants in all iterations.

The next step is how to analyze the converged pheromone matrix got by EPA in order to attain the final clustering solution of the network. Because of the convergence property of ant colony optimization, the community characteristic shown by this matrix is very obvious. Thus, to analyze this matrix, a simple partition method is adopted here, so that the community structure of the network will be naturally attained. The description of the partition phase algorithm (PPA) is given as follows.

Procedure PPA
Input: $B$ /* pheromone matrix after algorithm converging */
Output: $labels$ /*final clustering solution, or called community structure */
Begin
1  $labels \leftarrow zeros(1, n)$;
2  For $i=1: n$
3   If $labels(i)==0$
4    $V \leftarrow B(i, :)$; /* get the i-th row of pheromone matrix B */
5    $community \leftarrow find(V > \varepsilon)$; /* attain the community of node i; ε is sum(V)/n */
6    $labels(community) \leftarrow i$;
7   End
8  End
End

Because of the convergence properties of the exploration phase algorithm EPA, we can develop such a simple method PPA as partition phase algorithm. After the EPA converges at last, in the pheromone matrix $B$, each community will mix together and the rows which correspond to the nodes in the same community will be equal. Thus, by choosing any row from $B$, we can identify a community by using the average value $\varepsilon$ as a cutoff value to divide this row.

**Proposition 3.** The time complexity of our algorithm MACO is $O(TSln^2)$, where $T$ is the limitation of iteration number, $S$ is the size of ant colony, $l$ is the step number of each ant, and $n$ is the numbers of nodes.

***Proof.*** It's obvious that, the most computationally costly step of our method MACO is the 6-*th* step in its EPA subroutine, which generates each ant's local community in each iteration. The time complexity of this step is $O((l+d)n^2)$ known from proposition1 and proposition2. As there are $T$ iterations and each iteration produces $S$ ants, thus the time complexity of the MACO will be $O(TS(l+d)n^2)$. Since the average degree $d$ in network $N$ is regarded as a constant because most complex networks are sparse graphs, thus the time complexity of the MACO can be also given as $O(TSln^2)$.

Please refer to "*Appendix B*" in order to observe the execution process of algorithm MACO. It's obvious that the EPA subroutine converges well in this example, and thus, the PPA subroutine can easily partition the converged pheromone matrix so as to attain the communities of this network.

**2.4. *Parameters setting***

There are four parameters: $T$, $S$, $\rho$, $l$ in this algorithm, which denote iteration number limitation, the size of ant colony, the updating rate of the pheromone matrix and the step number limitation, respectively. The first three parameters can be easily determined, which is set at $T = 20$, $S = 100$ and $\rho = 0.6$ according to our experience and some experiment results. However, a suitable setting of parameter $l$ is much more important and difficult. Here we give some qualitative analysis on the step number $l$, and later we will also offer some quantitative analysis on this parameter in the experimental section.

Obviously, the value of $l$ should not be too small because walking only for a few steps will prevent the ant from properly exploring the network. One reasonable choice for $l$ can be made with the aid of the average distance of a network. The distance between two nodes refers to the number of links along the shortest path connecting them. As most social networks are small-world networks, the average distance between two nodes was shown to be around 6, according to the theory of "six degrees of separation" [26]. For scale-free networks, the average distance is usually small too. The World Wide Web is one of the biggest scale-free networks that we have found so far. However, the average distance of such a huge network is actually about 19 [27]; that is, we can get to anywhere we want through 19 clicks on the average. Thus, based on the above general observations, a good option for the value of $l$ that we propose should be $l = 20$.

It is noteworthy that, according to the parameters setting above, here the time complexity of our algorithm MACO can be also given by $O(n^2)$.

**3. Experiments**

In order to evaluate the performance of our algorithm MACO, we tested it on benchmark computer-generated networks as well as some widely used real-word networks. We conclude by analyzing the step number parameter $l$, which is defined in this algorithm.

In the experiment, our MACO is compared with four representative community detection algorithms, in which FN [5] and FUA [7] are optimization based method, while FEC [19] and LPA [11] are heuristic based method. Note that algorithm FUA has been regarded as one of the most effective method for community detection by the survey of Fortunato [28].

All experiments are done on a single Dell Server (Intel(R) Xeon(R) CPU 5130 @ 2.00GHz 2.00GHz processor with 4Gbytes of main memory), and the source code of all the algorithms used here can be obtained from [29].

**3.1. *Computer-generated networks***

We adopt two kinds of randomly-generated synthetic networks (by both Newman model [3] and Lancichinetti model [22]) with a known community structure to evaluate the performance of the algorithms. Moreover, here we employ a widely used accuracy measure so called Normalized Mutual Information (NMI) [30]. The NMI measure which makes use of information theory is regarded as a more fair metric compared with the other ones [30].

The first type of synthetic networks employed here is that proposed by Newman at al. [3]. For this benchmark, each graph consists of $n = 128$ vertices divided into 4 groups of 32 nodes. Each vertex has on average $z_{in}$ edges connecting it to members of the same group and $z_{out}$ edges to members of other groups, with $z_{in}$ and $z_{out}$ chosen such that the total expected degree $z_{in}+z_{out} = 16$, in this case. As $z_{out}$ is increased from the small initial values, the resulting graphs pose greater and greater challenges to the community detection algorithms. In figure 1(a), we show the NMI accuracy attained by each algorithm as a function of $z_{out}$. As we can see, our algorithm MACO outperforms all the other four methods in terms of NMI accuracy on this benchmark.

In order to further evaluate the accuracy of these algorithms, a new type of benchmark proposed by Lancichinetti et al. [22] is also adopted here. It has some properties such as the heterogeneous distributions of degree and community size which has been found in most real networks. Like the experiment designed by Lancichinetti et al. in [22], the parameters setting for the Lancichinetti benchmark is as follows. The network size $n$ is 1000, the average degree $d$ is 15, the maximum degree $d_{max}$ is 50, the exponent for the degree distribution $\tau_1$ is 2, the exponent for the community size distribution $\tau_2$ is 1, the minimum community size $c_{min}$ is 20, and the maximum community size $c_{max}$ is 50; while the mixing parameter $\mu$ (each vertex shares a fraction $\mu$ of its edges with vertices in other communities) varies from 0 to 0.6 with interval 0.05. In figure 1(b), we show the NMI accuracy attained by each algorithm as a function of the mixing parameter $\mu$. As we can see, our algorithm MACO is just a bit worse than the FUA, while it outperforms the other three methods in terms of NMI accuracy on this more challenging benchmark.

Computing speed is another very important criterion to evaluate the performance of an algorithm. Time complexity analysis for the MACO has been given by proposition3 and refined in Sec. 2.4. Nevertheless, here we show the actual running time of the MACO from an experimental angle in order to further evaluate its efficiency. Here we also adopt synthetic networks based on Newman model [3]. For this application, each graph consists of $n = 100C$ vertices divided into $C$ groups of 100 nodes. Each vertex has on average $z_{in} = 10$ edges connecting it to members of the same group and $z_{out} = 6$ edges to members of other groups. The only difference between the networks used here and the former ones is that, now $z_{out}$ is fixed while the communities number $C$ is changeable. Figure 1(c) shows the actual running time of the MACO. Figure 1(d) show the square root of the running time by MACO, which is nearly proportional to the number of nodes in the network. It's obvious that, the experiment can validate the correctness of the analysis on the time complexity for MACO, which is $O(n^2)$ when the parameters are all set at constant.

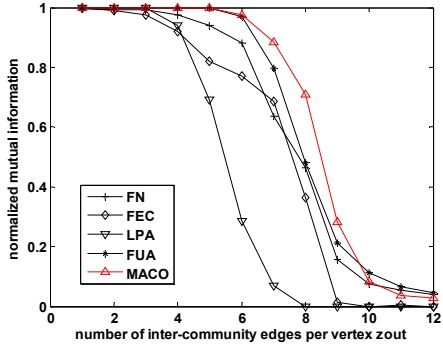

(a)

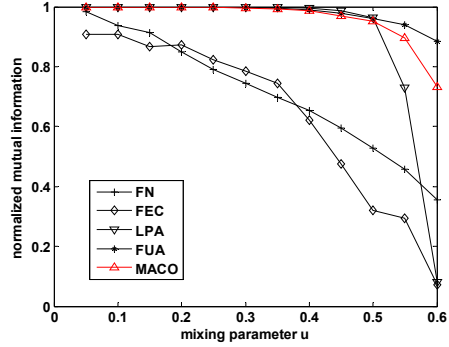

(b)

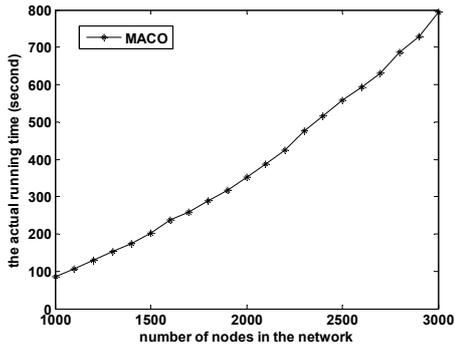

(c)

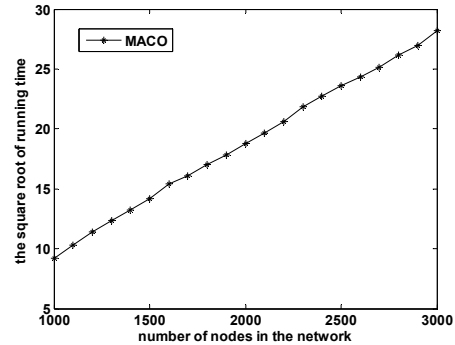

(d)

**Fig. 1.** (Color online) Test the performance of the MACO on artificial networks by both Newman model and Lancichinetti model. Each point is an average result over 50 graphs. (a) Compare MACO with FN, FEC, LPA and FUA in terms of NMI accuracy on Newman benchmark. (b) Compare MACO with FN, FEC, LPA and FUA in terms of NMI accuracy on Lancichinetti benchmark. (c) The actual running time of the MACO as a function of the network scale. (d) The square root of the running time by the MACO as a function of the network scale.

### 3.2. Real-world networks

As real networks may have some different topological properties from the synthetic ones, here we adopt several widely used real-world networks to further evaluate the performance of these algorithms. These networks that we use, and their sources and sizes, are listed in Table 1.

**Table 1.** Real-world networks used here.

| Networks | $|V|$ | $|E|$ | Descriptions |
|---|---|---|---|
| karate | 34 | 78 | Zachary's karate club [31] |
| dolphin | 62 | 160 | Dolphin social network [32] |
| polbooks | 105 | 441 | Books about US politics [33] |
| football | 115 | 613 | American College football [3] |
| jazz | 198 | 2,742 | Jazz musicians network [34] |
| email | 1,133 | 5,451 | Email network of human interactions [35] |

Because the inherent community structure for real networks is usually unknown, here we adopt the most commonly used network modularity function ($Q$) [36] to evaluate the performance of these algorithms. Table 2 shows the average result (over 50 runs) that compares our method MACO with FN, FEC, LPA and FUA in terms of function $Q$ on the real-world networks described in Table 1. As we can see, the clustering quality of our method MACO is competitive with that of the FUA, and better than that of the other three algorithms.

**Table 2.** Compare MACO with FN, FEC, LPA and FUA in terms of function $Q$ on real networks.

| $Q$-value | karate | dolphin | polbooks | football | jazz | email |
|---|---|---|---|---|---|---|
| FN | 0.3807 | 0.5104 | **0.5020** | 0.5497 | 0.4389 | 0.5037 |
| FEC | 0.3744 | 0.4976 | 0.4904 | 0.5697 | **0.4440** | 0.5173 |
| LPA | 0.3646 | 0.4802 | **0.5006** | 0.5865 | 0.3422 | 0.3706 |
| FUA | **0.4188** | **0.5268** | 0.4986 | **0.6046** | 0.4431 | **0.5406** |
| MACO | **0.4188** | 0.5081 | **0.5047** | 0.5917 | 0.4409 | 0.5490 |

### 3.3. Parameters analysis

Step number $l$ is a very important parameter in algorithm MACO, especially in its UC subroutine. Sec. 2.4 has given a reasonable indication on the choice of step number $l$. Here we also offer some quantitative analysis on this parameter from experimental angle.

In the UC, after each step, the transition probability vector $\psi_s^l$ will be updated and, thus, the ranking of all nodes will be changed according to the probability values of arriving at them from source node $s$. As long as $\psi_s^l$ is stationary, or say all members of the source community are put on the top of the sorted nodes sequence, it will be good enough for our purpose to unfold a community.

Thus, the convergence of UC can be evaluated based on the convergence of the transition probability vector, or that of the sorted node sequence. Fig. 2 shows the convergence process of the UC with the increase of step number $l$. In Fig. 2(a), the $x$-axis refers to the $l$-value, and the $y$-axis refers to the difference between the two consecutive transition probability vectors, which is defined as the Euclidean distance between them. In Fig. 2(b), the $x$-axis still refers to the $l$-value, while the $y$-axis refers to the difference between the two consecutive sorted nodes lists $L_1$ and $L_2$, defined as nnz($L_1$–$L_2$), which counts the number of non-zeros of a given vector.

We have tested different $l$ values in the range of $1 \leq l \leq 50$ for all real networks mentioned in the paper. It's obvious that the transition probability vector and the sorted nodes list can both converge well within 20 steps on each of the networks, and the UC's performance is insensitive to the choice of the parameter $l$ when it is greater than 20.

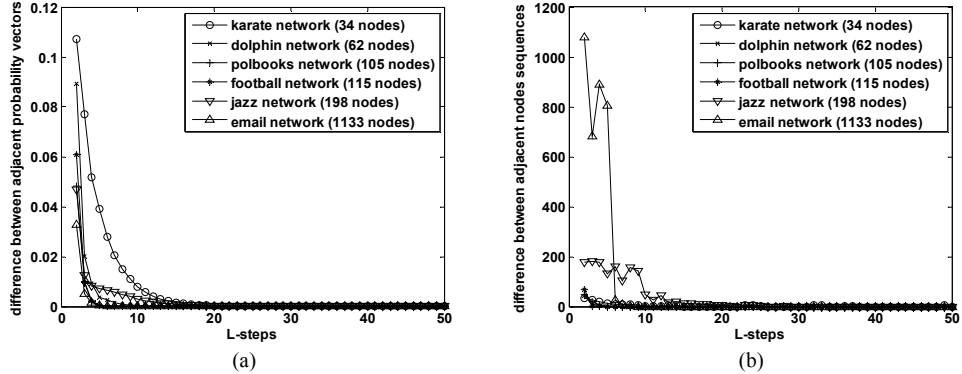

**Fig. 2.** (Color online) The sensitivity analysis of parameter $l$ on all the real networks used in this paper. For each network, the node with maximum degree is selected as source node $s$ here. (a) The convergence process of the transition probability vector with the increase of step number $l$. (b) The convergence process of the sorted node sequence with the increase of step number $l$.

## 4. Conclusion

A community detection algorithm MACO is proposed in this paper. The framework of ant colony optimization is employed as its basic algorithm framework. In each iteration, each ant detects its own community by a new Markov method with the aid of pheromone; all of the ants' local solutions are aggregated to a global one through a thought of clustering ensemble, which then will be used to update the pheromone matrix. As iteration proceeds and gathered knowledge on the past walks is registered on the network, the pheromone on within-community links becomes thicker, and that on between-community links becomes thinner. This process makes the ants' movement decision become more and more intelligent, and the trend that any ant remains in its own community becomes increasingly obvious. Once the algorithm converges, the community structure of the network will be obtained naturally from the pheromone distribution.

The main contribution of this paper is to propose a community detection algorithm MACO which possesses high clustering quality, while its efficiency is still not ideal to deal with some large-scale networks, such as WWW, Internet etc. Thus, the efficiency of our MACO should be further improved by exploring some potential optimization angles. Moreover, in our future work we intend to apply our method MACO in some interesting research areas, such as biological networks analysis, Web community mining, etc., and try to uncover and interpret the significative community structure that is expected to be found on them.

**Acknowledgment**

This work was supported by National Natural Science Foundation of China under Grant Nos. 60873149, 60973088, the National High-Tech Research and Development Plan of China under Grant No. 2006AA10Z245, the Open Project Program of the National Laboratory of Pattern Recognition, and the Erasmus Mundus Project of European Commission.

**Appendix A. The process that an arbitrary ant produces its solution in each iteration.**

In order to illustrate the process by which each ant produces its solution in each iteration within algorithm MACO, a Newman benchmark network [3] is used here. This graph contains $n = 128$ vertices which divided into 4 groups of 32 nodes. Its expected within-community degree of each node is $z_{in} = 9$, and its expected between-community degree of each node is $z_{out} = 7$. It's obvious that the community structure of this network is a little ambiguous. Without loss of generality, we always select the ant whose source node is 1 here. For the ant, we record three groups of results in each iteration. These results include: "the $l$-step transition probability vector $\psi$", "the cutoff position corresponding to the minimum conductance", and "this ant's local solution". A detailed description of them is given by Fig. A.1. Because of limited space, this figure only shows the results at the 1st generation, the 5th generation, the 10th generation, the 15th generation and the 20th generation, and denote them by using (a), (b), (c), (d) and (e) respectively. As we can see, our UC subroutine can clearly unfold the ant's local community, and the EC subroutine is effective to extract this community. Moreover, with the time passing, which means the pheromone on within-community links becomes thicker and thicker and the pheromone on between-community links becomes thinner and thinner, the local community detected by this ant becomes more and more precise.

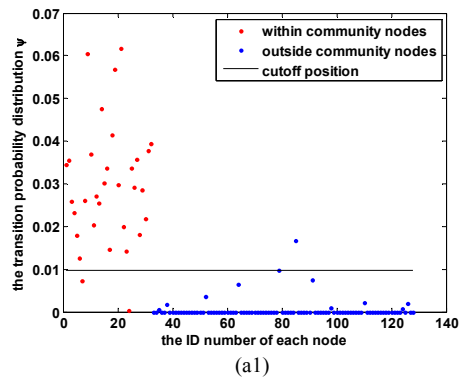
(a1)

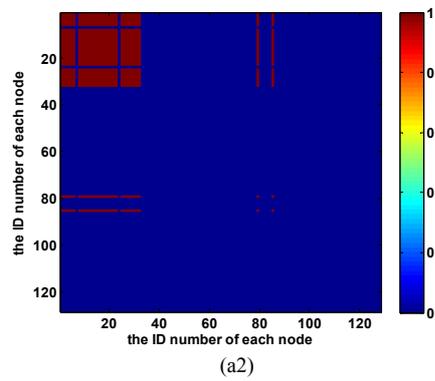
(a2)

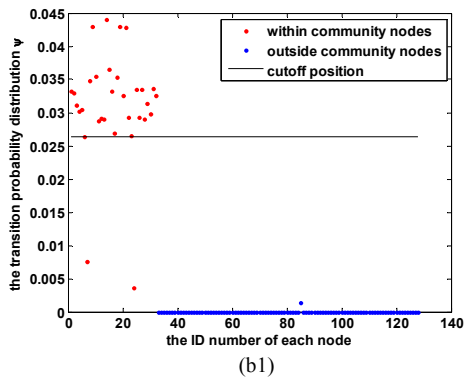
(b1)

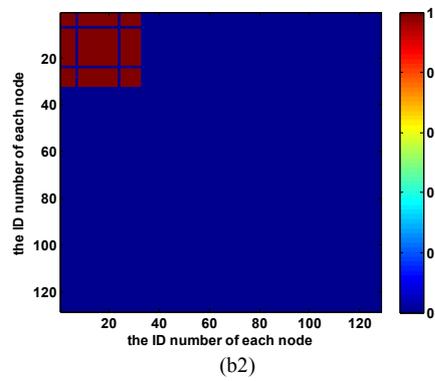
(b2)

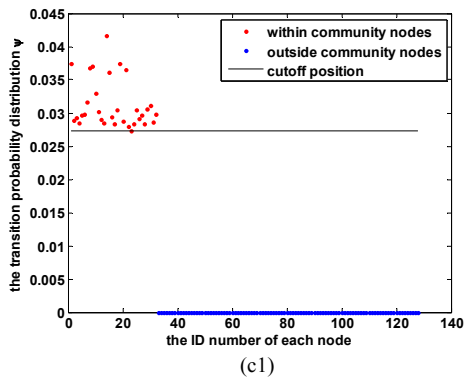
(c1)

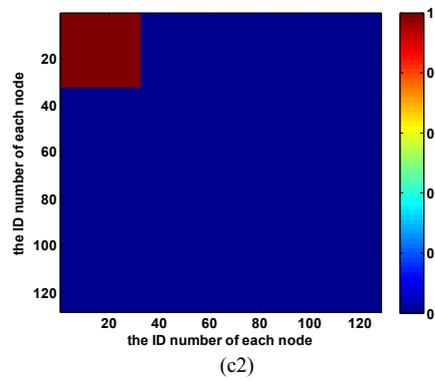
(c2)

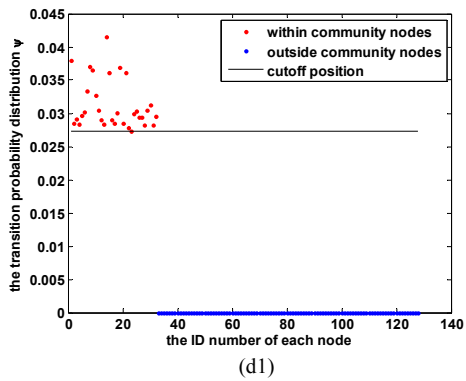
(d1)

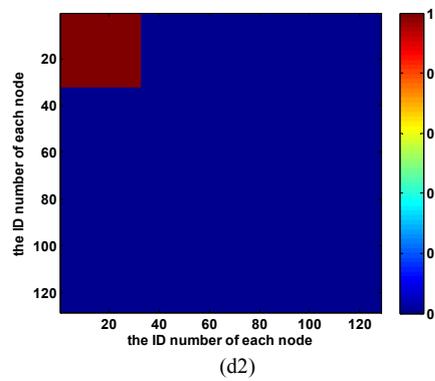
(d2)

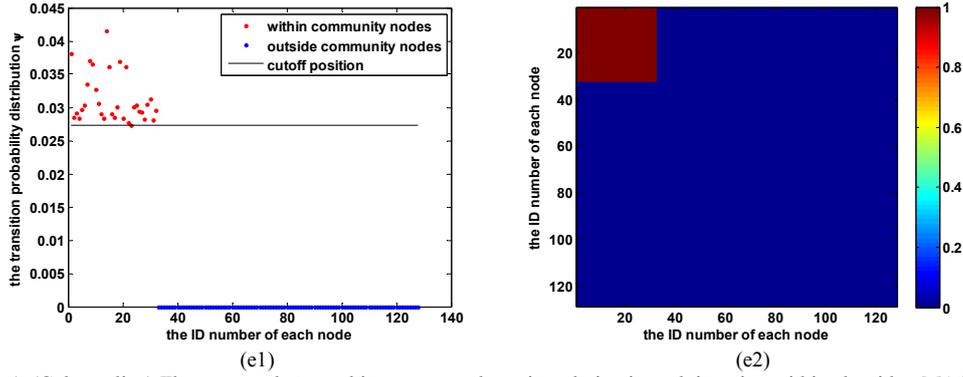

**Fig. A.1.** (Color online) The process that an arbitrary ant produces its solution in each iteration within algorithm MACO. Figures (a1)-(e1) show the ant's *l*-step transition probability vector and the corresponding cutoff position at the 1st generation, the 5th generation, the 10th generation, the 15th generation and the 20th generation respectively. Figures (a2)-(e2) show the ant's local solution at the 1st generation, the 5th generation, the 10th generation, the 15th generation and the 20th generation respectively.

## Appendix B. The execution process of algorithm MACO.

Again, in order to illustrate the execution process of our method MACO, and similar to the experiment conducted in *Appendix A*, we record two groups of results in each iteration. These results include: "the current global solution which aggregates the local solutions of all ants", and "the current pheromone matrix". A detailed description of them is given by Fig. B.1. This figure shows the results at the 1st generation, the 5th generation, the 10th generation, the 15th generation and the 20th generation, and denote them by using (a), (b), (c), (d) and (e) respectively. As we can see, our EPA subroutine can converge very well in this example, and thus, the PPA subroutine can easily partition the converged pheromone matrix so as to attain the communities of this network.

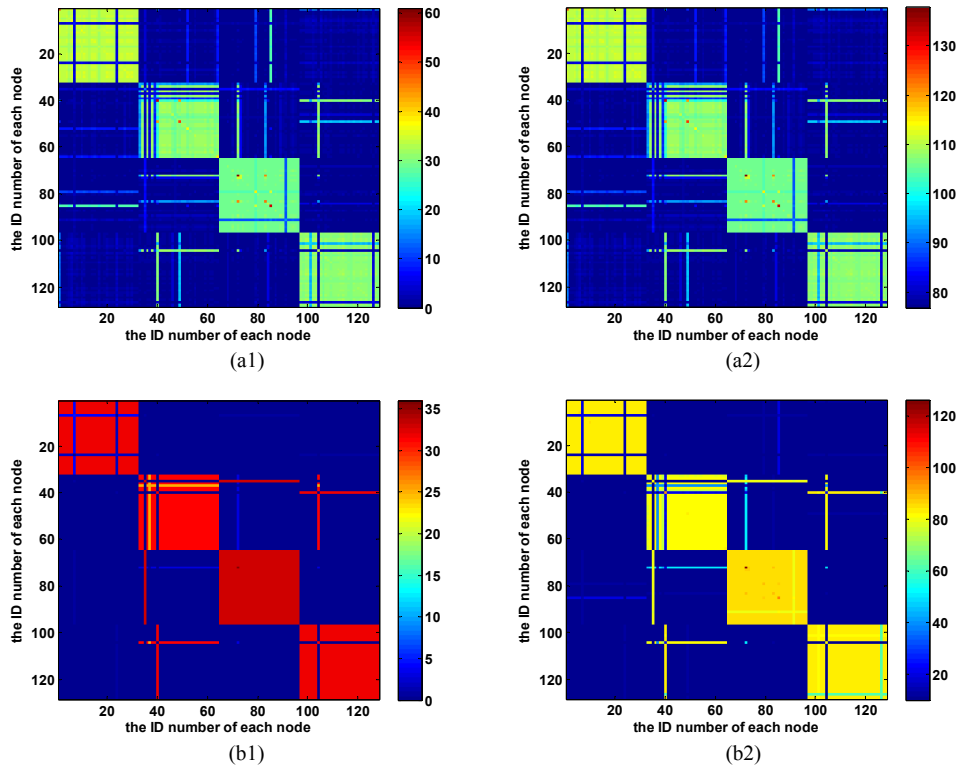

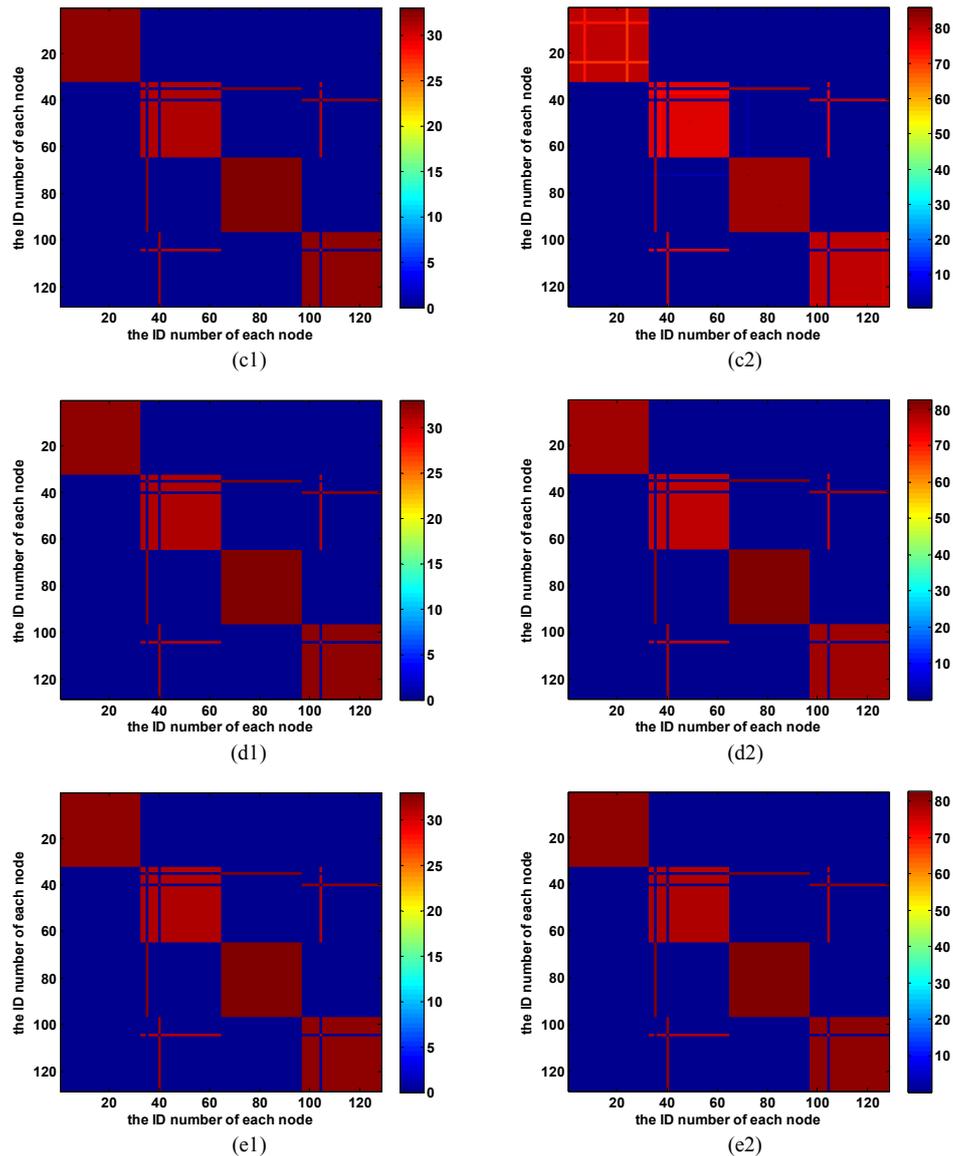

**Fig. B.1.** (Color online) A run of algorithm MACO. Note that Fig. A.1 and Fig. B.1 comes from a same run of the MACO. Figures (a1)-(e1) show the aggregated global solution at the 1st generation, the 5th generation, the 10th generation, the 15th generation and the 20th generation respectively. Figures (a2)-(e2) show the pheromone matrix at the 1st generation, the 5th generation, the 10th generation, the 15th generation and the 20th generation respectively.